\documentclass[journal]{IEEEtran}

\ifCLASSINFOpdf
\else
\fi
\usepackage{amsmath}
\usepackage{graphicx}
\usepackage{graphicx}
\usepackage{xcolor}
\usepackage[bottom]{footmisc}
\usepackage{epstopdf}
\usepackage[left=0.40in,right=0.40in,top=0.45in,bottom=0.45in, textheight=18in]{geometry}
\usepackage{algorithm}
\usepackage[noend]{algpseudocode}

\makeatletter
\def\BState{\State\hskip-\ALG@thistlm}
\makeatother

\begin{document}
\raggedbottom
%
\title{Managing Service-Heterogeneity using Osmotic Computing}
%
%
%

\author{\IEEEauthorblockN{Vishal Sharma\IEEEauthorrefmark{1}, Kathiravan Srinivasan\IEEEauthorrefmark{2}, Dushantha Nalin K. Jayakody\IEEEauthorrefmark{5}, Omer Rana\IEEEauthorrefmark{3}, Ravinder Kumar\IEEEauthorrefmark{4}} \\
    \IEEEauthorblockA{\IEEEauthorrefmark{1}Department of Information Security Engineering, Soonchunhyang University, Asan-si 31538, Republic of Korea \\
        Email: vishal\_sharma2012@hotmail.com} \\
    \IEEEauthorblockA{\IEEEauthorrefmark{2}Department of Computer Science and Information Engineering, National Ilan University, Yilan County, Taiwan (R.O.C) \\
        Email: kathiravan@niu.edu.tw} \\
         \IEEEauthorblockA{\IEEEauthorrefmark{5}Department of Software Engineering, Institute of Cybernetics, National Research Tomsk Polytechnic University, Russia\\
        Email: nalin.jayakody@ieee.org }
        
    \IEEEauthorblockA{\IEEEauthorrefmark{3}School of Computer Science and Informatics, Cardiff University, Cardiff, UK \\
        Email: ranaof@cardiff.ac.uk} \\
    \IEEEauthorblockA{\IEEEauthorrefmark{4}Computer Science and Engineering Department, Thapar University, Patiala, Punjab, India \\
        Email: ravinder@thapar.edu}\\
   
}

\maketitle

\begin{abstract}
Computational resource provisioning that is closer to a user is becoming increasingly important, with a rise in the number of devices making continuous service requests and with the significant recent take up of latency-sensitive applications, such as streaming and real-time data processing. Fog computing provides a solution to such types of applications by bridging the gap between the user and public/private cloud infrastructure via the inclusion of a ``fog" layer. Such approach is capable of reducing the overall processing latency, but the issues of redundancy, cost-effectiveness in utilizing such computing infrastructure and handling services on the basis of a difference in their characteristics remain. This difference in characteristics of services because of variations in the requirement of computational resources and processes is termed as service heterogeneity. A potential solution to these issues is the use of Osmotic Computing -- a recently introduced paradigm that allows division of services on the basis of their resource usage, based on parameters such as energy, load, processing time on a data center vs. a network edge resource. Service provisioning can then be divided across different layers of a computational infrastructure, from edge devices, in-transit nodes, and a data center, and supported through an Osmotic software layer. In this paper, a fitness-based Osmosis algorithm is proposed to provide support for osmotic computing by making more effective use of existing Fog server resources. The proposed approach is capable of efficiently distributing and allocating services by following the principle of osmosis. The results are presented using numerical simulations demonstrating gains in terms of lower allocation time and a higher probability of services being handled with high resource utilization.
\end{abstract}


%
\IEEEpeerreviewmaketitle

\section{Introduction}
The advent of hybrid networks and service-oriented architectures lead to data processing models that can operate at different levels of the systems stack -- within a data center, at the network edge, and within-network (based on increasing availability of programmable network elements, such as routers and switches). The combination of Internet of things (IoT), cloud and fog computing lead to mechanisms to manage and support complex networks~\cite{satyanarayanan2015edge}. Nowadays, each handheld user device is expected to synchronize their state over a  cloud-based data center, since large data and processing is involved which cannot be handled by the device itself (often due to battery life constraints and limited storage/processing capability). Consequently, with such ``mobile offloading'' based approaches, a large amount of data is expected to be transferred over the internet either to a private or public cloud for processing, evaluation, and storage.

With the significant increase in the type and range of {\it gadgets}, the heterogeneity of data has increased, which in turn has increased the types of services that need to be handled. Such services can carry out a range of different functions, e.g.: (i) data format conversion (between different vendors/manufacturers), (ii) encryption/decryption support; (iii) data sampling and aggregation/fusion across different devices, etc. Services can, therefore, differ in their computational and storage requirements. Although data classification has always been there to understand and evaluate the information out of it, but with a huge amount, the data is transferred to off-site servers for faster computations as well as storage. The retrieval and presentation are the other tasks performed over the data. Cloud computing has already changed the mechanism of data storage, retrieval, processing, and presentation.

Further, the issues related to the requirement of high speed as well as long-distance data transmissions are handled using the concept of fog computing, which allows the formation of a near-site private cloud. Fog computing reduces the computations and latency involved in the retrieval and processing of data over the public/private cloud.

Fog computing has undoubtedly provided a solution to reduce application latency, by making use of edge computing resources, but has raised the issue of a cost involved in making use of yet another cloud system~\cite{bonomi2012fog}~\cite{bonomi2014fog}~\cite{stojmenovic2014fog}. This cost is bearable where latency is a major constraint for the quality of experience for a user, but in general, this may increase the redundancy of available resources (thereby also reducing resource utilization) by making available a fog server for the same task as a server at an existing public cloud. Hence, the additional resources made available through a fog computing system is expecting to act as a surrogate for a similar resource that would (traditionally) be hosted within a cloud data center. Apart from these, complex service handling and heterogeneity are other issues which are to be handled in these cloud-fog architectures~\cite{lien2016collaborative}~\cite{lipowski2012roulette}~\cite{corcoran2016mobile}. Complex service handling involves authentication mechanisms, evaluation of complex data structures, and header evaluations of data from different sources.

Service heterogeneity arises because of a difference in the characteristics of operations and processes~\cite{sanaei2014heterogeneity}. Consider a server which receives continuous requests to handle some services that require consistent processing and involve complex data evaluations. Further, the same server receives similar requests from other sources with lesser computations, but in bulk. Now, this server is under a tremendous pressure of executing two different types of services, which in total have the same impact on its operations. However, one of these services is executed as a single unit, while other is executed as a batch. This creates variety in the type of services being handled by a server resulting into large heterogeneity, which is difficult to manage and control during operations. Thus, a division of these services allows allocation of appropriate servers, which are best suited to handle them resulting in the formation of an efficient processing environment. Management of service heterogeneity is complex, but it can be attained by dividing processing on the basis of micro- and macro-services.

The definition of micro- and macro-services may vary from scenario to scenario, but in this paper, this classification is performed on the basis of requirement of computational resources, energy consumption and processing time. The division of services on the basis of classification and allocation of appropriate computational resources remains a challenge. Efficient resource provisioning can resolve these issues in fog computing by providing delay-sensitive systems~\cite{skarlat2016resource}. Despite this, a broader solution is required, which can operate in any environment irrespective of the classification as well as the categorization of services and available resources. An efficient solution for handling of services can be obtained by the utilization of a new paradigm of computing as stated by Villari et al.~\cite{7802525} and termed as ``Osmotic Computing". This paradigm advocates the need for movement of microservices between a data center and edge devices to support load balancing. It also reduces the latency of the overall application but does not fully identify mechanisms that can be used to support and facilitate such service migration.

\begin{figure*}[!ht]
  \centering
  \includegraphics[width=300px]{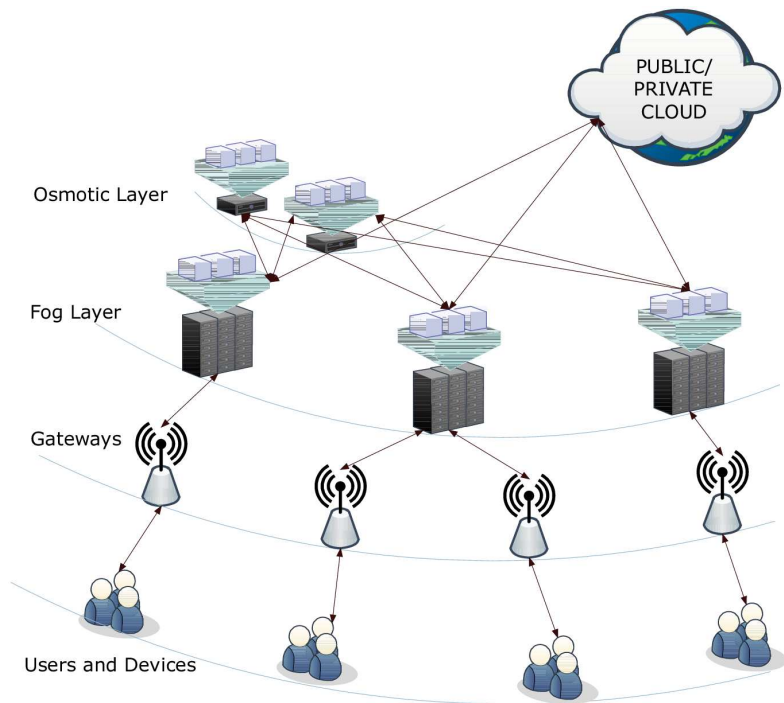}
  \caption{An illustration of operational view of layers in osmotic computing.}\label{second}
\end{figure*}
With an enhancement in the capability, capacity, processing, divisibility and reduction in latency and overheads, osmotic computing can provide better resource utilization within a cloud and fog computing environment. Osmotic computing is aimed as a mechanism to support service migration across cloud, fog and edge computing resources. This paper utilizes fog-based resources as an integral part to implement osmotic computing, as shown in Fig.~\ref{second}. Osmotic computing is derived from the term ``osmosis" which refers to the equalization of the concentration of a solution by allowing the solvent to move through a semipermeable membrane. Here, concentration refers to the mixture of solute and solvent. A similar analogy can be applied to modern day computing infrastructure, by separating the services to acquire processors with an aim of balancing the load as well as resource utilization without any redundancy. Although, the initial concept of osmotic computing is to balance the services across the servers, but it can also be used to simply migrate the services across different computing platforms.

In this paper, the concept of osmotic computing is studied with its implementation for controlling and managing service heterogeneity which arises due to a large number of services being handled by the same server. The primary focus of the paper is in the selection of resources (within a computational infrastructure) that contribute to a ``solute" and ``solvent", and form the key part of the ``solution" (taking the osmosis analogy). Theoretical and numerical analyses are presented which allow understanding the advantages and impact of osmotic computing within the context of a particular application deployment. A fitness-based osmosis algorithm is proposed which controls the movement of services between the osmotic layers and public/private cloud layers.

\section{Motivation and Problem Statement}
\label{sec:motivation}

Osmotic computing is presented as a new paradigm by Villari et al.~\cite{7802525}, which can reduce the gap between the use of edge devices (i.e. devices directly owned and operated by a user) and the cloud-based data centers. What constitutes as ``edge" resources can vary from one application to another. Some consider the immediately managed device, such as a handheld smartphone or a surveillance camera as an edge resource, others also take account of the first hop network component (e.g a router or a switch) as part of the overall edge resource. The proposed osmotic computing paradigm attempts to specify the efficient handling of available resources by improving the capability as well as the reach of the servers' computational power closer to the user. With the advent of near-site osmotic computing, it is necessary to study and analyze the possibilities of implementing an infrastructure for osmosis. Identifying how microservices can be migrated from edge resources to cloud-based resources (and vice versa), and characteristics which influence such migration,  remains a challenge -- a requirement that has not been fully articulated in the osmotic computing vision in Villari et al.~\cite{7802525}.

Considering computational infrastructure as a chemical solution whose properties can change over time, our focus is on identifying the properties of what constitutes a solute and solvent, which is then operated on the principle of osmosis to manage and control services. We also take into account the need to maintain some core services (which do not migrate) -- such as resource monitoring, security support, etc, as outlined in Villari et al.~\cite{7802525}. Further, with a large number of services being handled by a single server, the classification of services and their distribution needs to be influenced by changes in resource properties and user application characteristics. Thus, an approach is required which can control and manage these heterogeneous services by efficiently allocating them across different servers.

\section{Proposed Approach}
\label{sec:proposedapproach}

The proposed approach considers a hybrid scenario comprising edge devices making multiple service requests to a cloud environment. The initial approach begins with the classification of the solute, solvent, and concentration. Next, a system model is formulated to derive a fitness function and an algorithm which influences the shift of services between the different layers for efficient control and management of services.

\subsection{Taxonomy}
Considering the definition of osmosis, there are five major pillars as explained below:
\begin{itemize}
  \item Solute: The soluble part which is not allowed through the semipermeable membrane is termed as the solute. In the proposed approach, computational power and energy, processing time, and current load are considered as the solute. However, these are not the only components; more properties can be considered as a part of solute depending on the scenario and need of the applications.
  \item Solvent: The part of the solution which absorbs the solute is termed as the solvent. This is the only component which is allowed to move through the semipermeable membrane. With a focus on the service-based scenario, the services are to be shifted between the servers by dividing them into micro- and macro-services. Thus, the services form the solvent part of the solution.
  \item Solution: The entire infrastructure, comprising users, servers, services, and resources, form the solution.
  \item Semipermeable membrane: The benefits of the approach depend on the selection of a semipermeable membrane which takes a decision about the movement of solvent across the solution to maintain appropriate concentration. In the proposed approach, fog servers are considered as the semipermeable membrane. The movement is based on considering: (i) when to \emph{move} -- i.e. when micro-services should be migrated from our type of resource to another; (ii) where to \emph{move} -- i.e. should micro-services be moved to a data center or to an edge resource; (iii) how to \emph{move} -- i.e. the mechanism used to support the migration -- this may involve deployment of lightweight containers, or aggregation of containers into pods (as in Kubernetes), etc.
  \item Concentration: The concentration represents the ratio of solute to solvent, i.e., in osmotic computing, it is considered as the ratio of the number of services to be handled by the total computational resources available at the server layer. An architecture can have one or more osmotic layers, depending on the complexity of services and number of users making requests.
\end{itemize}

\subsection{Network Architecture}
\label{sec:networkarch}

Although osmotic computing aims at bridging the gap between the edge cloud and public cloud, a similar approach is also demonstrated using the fog computing, however the later is primarily focused on the need to provision ``cloudlets'' in close proximity to a user that itself requires an efficient solution for deciding the procedures for service execution. Fog servers/cloudlets lead to enhancement in the speed of processing of an application by reducing the latency involved in the movement of data between the users and the public cloud. Osmotic computing also focuses on the similar issue of latency, but with a primary target on the control of services by dividing them into micro- and macro-components, and then migrating them to an infrastructure for implementation of non-redundant and faster computations.

The proposed approach considers the impact of fog computing and focuses on the implementation of osmotic computing via fog servers. The osmotic layer in the proposed approach, therefore, makes use of fog servers as the decision point, in deciding whether services should be executed on a fog server or migrated to a cloud-based data center. The architecture can be modified on the basis of available resources and implementation requirements.

In the proposed solution, a four-layered model is used for handling the services of users without much redundancy as well as latency, as shown in Fig.~\ref{second}. The initial layer is the user plane which interacts with the servers for the handling of their service requests. The interacting servers are the second layer which is the actual fog computing layer containing simple processing server which acts as the semipermeable membrane that takes a decision on the classification of services into micro- and macro-components. This layer is connected to two layers, namely, osmotic layer which contains servers that can perform computations over micro-services and public cloud layer for handling the large service requests.

The heterogeneous services can be easily moved between the osmotic and public/private cloud layer to allow efficient control over each service. This allows easy identification of the servers which control a particular type of service. Thus, making it efficient to know entire information without much latency as well as redundancy. The proposed model utilizes the osmosis principle which is modeled over service and computational sets as presented below.

Let $S$ be the set representing the total services generated from the user layer with $N$ being the set of resources available at osmotic layer and the $M$ being the number of resources reserved at the public cloud. The fog servers can vary in number, depending on the deployment scenario being considered. The fog serves can also operate as the osmotic servers to handle micro-services. The set $S$ is subdivided by the fog servers into two subsets, namely $S_{a}$ and $S_{b}$ representing micro and macro-services, respectively. The division is based on the principle of equalizing the concentration $C$ which is given as:
\begin{equation}\label{eq:1}
  C=\frac{|S|}{|R|}
\end{equation}
where $R$ is the set of resource properties that includes load over the servers ($L$), total energy consumption ($E$) and resource reservation time ($\tau$). The expression in (\ref{eq:1}) can be used for trivial applications comprising scenarios with one-to-one service requests; however, for other types, a fitness function $f_x$ is used to model the concentration which is given as:
\begin{equation}\label{eq:2}
  f_x=\frac{\alpha_1 R_{1}+ \alpha_2 R_{2}+ \dots + \alpha_k R_{k}}{\alpha_1+\alpha_2+\dots + \alpha_k}, R_{i \dots k} \in R, \;k=|R|.
\end{equation}
Here, $\alpha$ represents the fitness weight of the property such that $0 \leq \alpha \leq  1$ and $\alpha$ can be treated as a dependent or independent variable. For dependent, $\sum_{i=1}^{k}\alpha_{i}=1$, whereas as for independent $\sum_{i=1}^{k}\alpha_{i}\neq 1$. The weight is provided on the basis of dominance of a resource property i.e. for dependent scenarios, the dominant property possess $\alpha=0.5$ whereas the other possess $\alpha=\frac{0.5}{k-1}$; for independent consideration, $\alpha$ is the ratio of allocated components to the total available components.

In the proposed approach, three major resource properties are considered, thus, (\ref{eq:2}) deduces to
\begin{equation}\label{eq:3}
  f_x=\frac{\alpha_1 L_r+ \alpha_2 E_r+  \alpha_3 \tau_r}{\alpha_1+\alpha_2+\alpha_3},
\end{equation}
where $L_r$, $E_r$ and $\tau_r$ are the load, energy and processing time for the requested services. Here,
\begin{equation}\label{eq:4}
  \alpha_1=\frac{L_a}{L_t},
\end{equation}
\begin{equation}\label{eq:5}
  \alpha_2=\frac{E_c}{E_t},
\end{equation}
and
\begin{equation}\label{eq:6}
  \alpha_3=\frac{\tau_p}{\tau_t},
\end{equation}
where, $L_{a}$ is the current load handled by the servers, $L_{t}$ is the total load the servers can handle, $E_{c}$ is the energy already consumed, $E_{t}$ is the total energy available, $\tau_p$ is the processing time consumed and $\tau_t$ is the total time available to retain/use a server. The processing time cannot go beyond the available time (often identified as a ``deadline" for latency sensitive applications). A shift to another resource cannot be made if any of the fitness weight ($\alpha$) acquires a value 1 (as this indicates that the user has indicated a preference for a particular type of resource). The operational activity of the proposed approach will be highly dependent on the distribution of users and services. In the proposed model, a service description that enables high flexibility (i.e. where values of $\alpha_{i} < 0.5$) enable better migration of services between the different available resources.

The fitness function forms the basis for the fog servers to take a decision on moving the services till the fitness of public and the osmotic clouds are not equal or within some upper or lower bound. The fog servers take a decision on the division of services as well as on the movement between the different layers of the proposed model. The division can be made by an incremental method which includes sub-division of services until their fitness value becomes equal to fitness value of available servers on either layer, or it can be undertaken by setting a threshold on the property of every incoming service request.

A service requiring resources greater than a certain threshold can be classified into a set $S_{b}$ and transferred to the public cloud, whereas the others are classified into a set $S_{a}$ to be handled by osmotic servers. The division allows management and control of the heterogeneous services on the basis of a fitness value. In the case of non-divisibility of services and failure to acquire a fitness value large enough to shift between the layers, various scheduling approaches can be used to prioritize the services for efficient handling.

After finalizing the system model, a fitness-based osmosis algorithm is used to transfer the services between the osmotic and cloud layers. This algorithm operates over the fog servers and is a mini-decision support system in itself. Currently, only a limited number of parameters are used just to demonstrate the impact of osmotic computing; however, in future studies, it is required to consider a fully-independent architecture with a focus on other aspects of cloud and edge resources.

\begin{algorithm}[!ht]
\fontsize{8}{10}\selectfont
\caption{Fitness-based Osmosis for Services}
\label{algo1}
\begin{algorithmic}[1]
\State \textbf{Input}: $N$, $S$, $M$, $R$
\State \textbf{Output}: $f_{x}^{osmotic}$ = $f_{x}^{public}$ $\pm \epsilon$, track
\State set $\epsilon$
\State set track=0
\State Evaluate Eqs. (\ref{eq:3})-(\ref{eq:6})
\State Calculate threshold $f_{x,th}^{osmotic}$ for osmotic layer
\State Calculate threshold $f_{x,th}^{public}$ for public cloud layer
\While {($f_{x}^{osmotic}$ $<$ $f_{x}^{public} \pm \epsilon $ )}
\If{($f_{x}^{service}$ $\leq$ $f_{x,th}^{osmotic}$)}
\State shift service to $S_{a}$
\State Recalculate $f_{x}^{osmotic}$
\ElsIf{($f_{x}^{service}$ $\geq$ $f_{x,th}^{osmotic}$) \&\& ($f_{x}^{service}$ $\leq$ $f_{x,th}^{public}$)}
\State shift service to $S_{b}$
\State Recalculate $f_{x}^{public}$
\Else
\State Message.out(``Cannot be handled currently")
\State adjust $\epsilon$ and reset
\EndIf
\State \textbf{end if}
\State track=track+1
\EndWhile
\State \textbf{end while}
\end{algorithmic}
\end{algorithm}

The steps for performing the initial levels of processing in osmotic computing are presented in Algorithm~\ref{algo1}. The algorithm takes into account the current infrastructure and service scenarios which are then evaluated over different fitness values until the concentration (fitness) of the entire network is not stabilized over equality. A varying limit $\epsilon$ is considered which allows the shift in fitness value to consider a service for osmotic or public cloud layer. The value of $\epsilon$ and $\alpha$ play a key role in shifting services between the two layers. This algorithm presents the basic osmosis approach which helps in managing the services on the basis of their heterogeneity and complexity in terms of requirement of computational resources. The algorithm operates till the fitness equalization is not obtained and gives ``track" as an outcome, which represents the number of iterations a system has to undergo for distributing services between the layers; it also forms the complexity of the proposed algorithm. In the worst case, every service is shifted and epsilon is updated in every iteration. Thus, the running time depends on the number of times the value of $\epsilon$ is updated and the services shifted. Since $\epsilon$ is updated as a procedure itself, the worst case complexity is of the order $O(|S|\times x)$, where $x$ is the number of times $\epsilon$ is updated.
\subsection{Theoretical Analyses}
This section evaluates the proposed osmotic computing approach over two paradigms, namely, probability of selecting a service and the effect on the number of iterations with variation in $\epsilon$.\\

\textbf{Theorem-1:} The probability of service being selected in every iteration depends on the fitness variation w.r.t. threshold value and is given as:
\begin{equation}\label{eq:7}
 \frac{1}{f_{x,th}^{public}} \leq \frac{1}{\sum f_{x}^{service}} \leq \frac{1}{f_{x,th}^{osmotic}}
\end{equation}

\textbf{Proof:} This can be explained with the help of Roulette-wheel selection~\cite{lipowski2012roulette}, according to which, the probability of being selected is given as the ratio of the fitness value of a service to the sum of fitness values of all services, i.e.
\begin{equation}\label{eq:8}
  P(s)=\frac{f_{x}^{service}}{\sum f_{x}^{service}}.
\end{equation}
Now, for the service being handled by the public cloud, the probability is given as:
\begin{equation}\label{eq:9}
  P(p,s)=\frac{f_{x}^{service}}{\sum f_{x,th}^{public}},
\end{equation}
and for the service being handled by the osmotic layer, the probability is given as:
\begin{equation}\label{eq:10}
 P(o,s)=\frac{f_{x}^{service}}{\sum f_{x,th}^{osmotic}}.
\end{equation}
For osmotic computing, $f_{x,th}^{public} \geq f_{x,th}^{osmotic}$, therefore $P(o,s) \geq P(p,s)$, which states that the probability of being handled by osmotic servers for a service with lower threshold is always higher irrespective of the resources, and by re-arranging, it can be concluded that it follows the following condition
\begin{equation}\label{eq:7}
  \frac{f_{x}^{service}}{f_{x,th}^{public}} \leq P(s) \leq \frac{f_{x}^{service}}{f_{x,th}^{osmotic}},
\end{equation}
 which satisfies the stated theorem.\\

\textbf{Theorem-2:} The value of $\epsilon$ affects the performance of the system and with a larger value of $\epsilon$, the number of iterations required to allocate the services decreases, which improves the performance.

\textbf{Proof:} In the proposed approach, $f_{x}^{osmotic}$ = $f_{x}^{public}$ $\pm \epsilon$ which is a condition for resource utilization in the entire network. Considering a larger value of $\epsilon$, the fitness function follows the rule of $f_{x}^{osmotic}$ $\leq $$f_{x}^{service}$ $\leq $ $f_{x}^{public}$ as $f_{x}^{service}$ satisfies the threshold conditions, which makes the requirement of extra iterations to be 0. This makes the entire procedure to be operated in a single iteration, i.e. $\epsilon=1$, which decreases the running time of the osmosis algorithm and it is of the order $O(|S|)$.
\begin{figure}[!ht]
  \centering
  \includegraphics[width=250px]{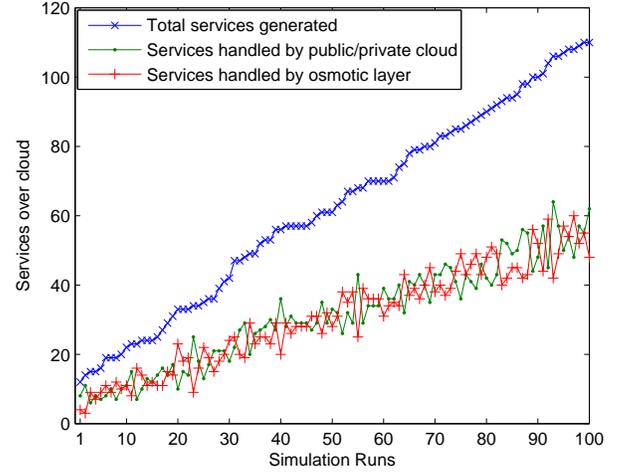}
  \caption{Service variations vs. simulation runs.}\label{g1}
\end{figure}
\begin{figure}[!ht]
  \centering
  \includegraphics[width=250px]{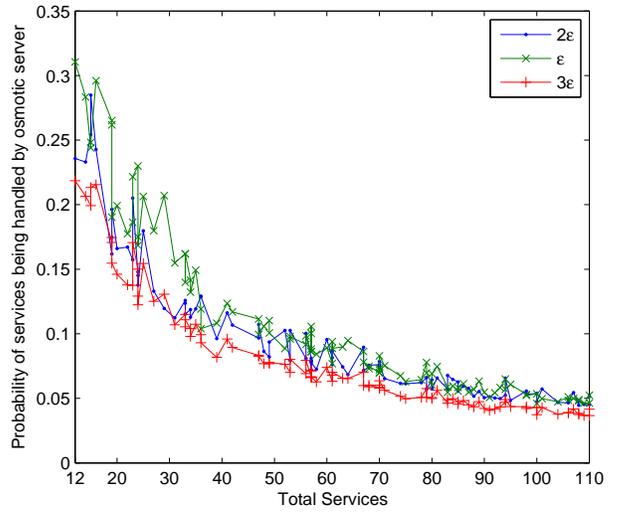}
  \caption{Probability of service handled by osmotic server vs. total services.}\label{g2}
\end{figure}
\begin{figure}[!ht]
  \centering
  \includegraphics[width=250px]{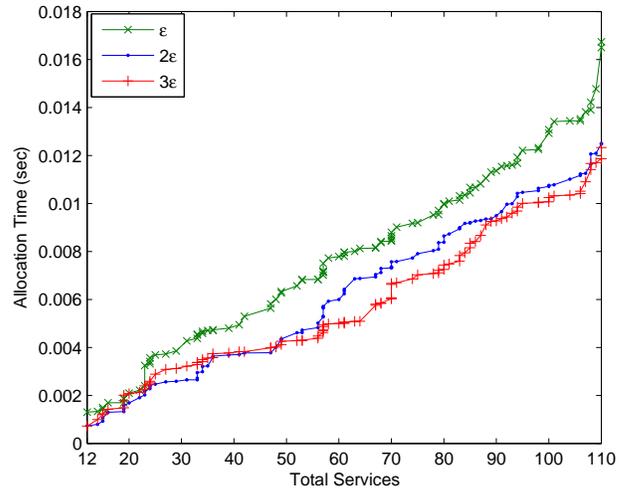}
  \caption{Server allocation time vs. total services.}\label{g3}
\end{figure}
\subsection{Performance Evaluation}
The proposed model and osmosis algorithm are tested using numerical simulations. An environment is considered which contains 12 to 110 service requests made by 10 users over an infrastructure comprising 10 public/private servers, 5 osmotic servers each capable of handling the load with a value of 10 concurrent requests. As stated earlier, evaluations are performed by dividing the services on the basis of the requirement for computational resources, energy consumption and processing time. The fog servers and osmotic servers are considered to be the same in this analysis. The total energy available with a single server is 2000 J; energy consumption per iteration is fixed at 1.5 J with a minimum processing time of 5 seconds and total available time of 100 seconds. The number of resource properties being considered is 3 (energy, load and time). Dependent environment is modeled with $\sum_{i=1}^{k}\alpha_k=1$. The initial value for $\epsilon$ is fixed at 100 and it is varied between $2\epsilon$ and $3\epsilon$ during analysis.

Initially, the results are recorded for the total number of services generated during each simulation run and the number of services handled by the osmotic and public/private cloud. Fig.~\ref{g1} presents the total number of services generated and their distribution between the osmotic layer and cloud layer. With osmotic layer predominating the near-site processing, a majority of the micro-services are processed by the osmotic servers, whereas the macro-services are handled by the large infrastructure support of the public/private cloud. Micro-services are only transferred to public/private cloud, when the osmotic servers are overloaded or when no adjustments are made to the value of $\epsilon$. Since the simulations are evaluated in a controlled environment, no adjustments and extra iterations are required to handle the services. All the services are classified and processed in a single iteration.

We also consider the probability of services being handled by the osmotic server with a variation in $\epsilon$ as shown in Fig.~\ref{g2}. The results show that with an increase in the value of $\epsilon$, the services are unable to satisfy the osmotic threshold, thus, the probability of being served by the osmotic server decreases. This may be due to a discrepancy between the resources at the osmotic layer, compared to those that are available within a data center/cloud. This decrease can be controlled by selecting an optimal value of $\epsilon$. Selection of $\epsilon$ and maintaining a value that is above threshold remain other open issues which need to be considered in further research.

Our theoretical analysis suggests that with an increase in the value of $\epsilon$, the number of iterations required to distribute/allocate the service should decrease. This is the main paradigm of service heterogeneity. A similar effect is observed during the numerical simulations as shown in Fig.~\ref{g3}. The time required to allocate the services to a particular server layer decreases with an increase in the value of $\epsilon$. The decrease is caused because of the dependency of threshold value over $\epsilon$. Considering the results in Figs.~\ref{g2} and~\ref{g3}, it can be identified that there is a tradeoff between the time of allocation and probability of being distributed to the osmotic server. It is noticed that with extra decisions involved in the allocation of services between the osmotic and public/private cloud, more time will be consumed, but this can reduce the latency involved in the processing of data as well as can prevent over consumption of resources.

From the results, it is evident that osmotic computing has laid the foundation of a new paradigm for computing and has objectified the scenarios for handling large data and computations on the basis of requirement and available resources.

\subsection{Open Issues}
\label{sec:openissues}

Osmotic computing provides an alternative perspective of migrating services between data center-hosted resources and resources that can be made available in closer proximity to a user (as advocated in fog computing). Emerging interest in making more effective use of increasing capability made available within such edge resources, as also observed in recent work in 5G networks, the proposed approach describes how three resource properties can be used to decide where a service should be executed.  With a focus on the service divisibility, there are several key challenges, which if taken care of can provide a fault-tolerant, robust,  low-latency approach for managing services. These include:

\begin{itemize}
  \item Resource scheduling: It is a key challenge in the case of osmotic computing. With a diversity in the type and range of services available and their allocation to different resources, scheduling plays a key role in the efficient handling and management of services. Scheduling in this instance can also be influenced by user mobility -- i.e. as a user migrates from one location to another, fog resources that may be suitable would also change. Understanding how such mobility can be taken into account in scheduling decisions is an important requirement to make more effective use of the osmotic computing paradigm.
  \item Resource allocation: With an introduction of the new fog computing/osmotic layer, it becomes necessary to understand how allocation can be supported across both local and remote resources. The allocation of resources on the basis of services and their classification is an important issue.
  \item Energy conservation: With the disposition of services, it is mandatory to prevent non-redundant allocation of services so as to preserve over-consumption of energy. This can be termed as ``Green Osmotic Computing".
  \item Aggregation and Distribution: Classification of services into micro and macro groups requires a standard, which is yet to be developed for such computing environment. The shift of services between the different resource layers depends on the successful implementation of aggregation and distribution strategies.
  \item Service migration: The key principle of osmotic computing is support for service migration, which is still an open issue and can be resolved considering various optimization solutions.
  \item Reverse osmosis as security: Security is a major concern for osmotic computing. Privacy and authentication are the major issues to be resolved for the successful implementation of osmotic computing. The principle of ``Reverse-Osmosis" can be considered to overcome the intrusions, where intruders are the impurities in the infrastructure as the solution.
\end{itemize}

\section{Conclusion}
\label{sec:conc}

Service provisioning and migration based on the osmotic computing paradigm proposed by Villari et al. (2016) is described. The approach advocated in this paper takes account of service requirements (based on three resource properties), to determine whether services should be executed at the data center or migrated to fog computing resources (in closer proximity to a user). We consider such service migration to be an important requirement as our edge computing infrastructure matures in capability.

With the aim of enhancing the capability and improving utilization of near-site computational infrastructure, we propose classifying services to enable better allocation of these to resources. Considering the principle and objective of osmotic computing, a migration algorithm is proposed in this paper, which utilizes a fitness function to distribute and allocate the services into micro- and macro-components. These components are handled by either an osmotic layer or a public/private infrastructure. The results are presented using numerical simulations, which states the importance of osmotic computing via proposed algorithm in terms of lower allocation time and a higher probability of services being handled without much utilization of resources.

Security remains an important challenge not fully considered in this work. Understanding how service migration can be supported across different layers of a computational infrastructure, subject to user privacy and infrastructure security capability, remains the next step in this work. One way to achieve this would be to support secure containers that host services and only enable migration of an entire container to a remote platform. Only fog computing resource which can host and deploy such a secure container are considered during the migration process. This could be included as an additional binary decision variable, beyond the three characteristics being considered in this work.

\bibliographystyle{ieeetr}
\bibliography{final_ref}








\end{document}